\documentclass[preprint,aps]{revtex4-1}
\usepackage{graphicx}
\usepackage{hyperref}
\usepackage{epstopdf}
\usepackage{slashed}
\usepackage{rotating}
\usepackage{float}
\newcommand{\be}{\begin{equation}}
\newcommand{\ee}{\end{equation}}
\newcommand{\bea}{\begin{eqnarray}}
\newcommand{\eea}{\end{eqnarray}}
\newcommand{\ba}{\begin{array}}
\newcommand{\ea}{\end{array}}

\begin{document}
\title{Magnetic moments of $J^P=\frac{3}{2}^+$ decuplet  baryons using effective quark masses in chiral constituent quark model}
\author{Aarti Girdhar}
 \affiliation{Department of Physics, Dr. B.R. Ambedkar National
Institute of Technology, Jalandhar, 144011, India \\ and Regional Centre for Accelerator-based Particle Physics,
Harish-Chandra Research Institute, Chatnaag Road,
Jhunsi, Allahabad 211019, India}
\author{Harleen Dahiya}
\affiliation{Department of Physics, Dr. B.R. Ambedkar National
Institute of Technology, Jalandhar, 144011, India}
\author{Monika Randhawa}
\affiliation{University Institute of Engineering and Technology,
Panjab University, Chandigarh, 160014, India}

\begin{abstract}

The magnetic moments of  $J^P=\frac{3}{2}^+$ decuplet baryons have been calculated in the chiral constituent quark model ($\chi$CQM) with explicit results for the contribution coming from the valence quark polarizations, sea quark polarizations, and their orbital angular momentum. Since the $J^P=\frac{3}{2}^+$ decuplet baryons have short lifetimes, the experimental information about them  is limited. The $\chi$CQM has important implications for chiral symmetry breaking as well as SU(3) symmetry breaking since it works in the region between the QCD confinement scale and the chiral symmetry breaking scale. The predictions in the model not only give a satisfactory fit when compared with the experimental data but also show improvement over the other models. The effect of the confinement on quark masses has also been discussed in detail and the results of $\chi$CQM are found to improve further with the inclusion of effective quark masses.
\end{abstract}
\maketitle
\section{Introduction}
\label{intro}
In order to understand the internal structure of the hadrons in the nonperturbative regime of Quantum Chromodynamics (QCD), one of the main challenges is the measurement of static and electromagnetic properties of hadrons like masses, magnetic moments, etc. both theoretically  and experimentally. Ever since the proton polarized structure function measurements in the deep
inelastic scattering (DIS) experiments \cite{emc,smc,adams,hermes} provided the first evidence that the valence quarks of proton carry only a small fraction of its spin, the charge, current and spin structure of the nucleon has been extensively studied in experiments measuring the electromagnetic form factors from the elastic scattering of electrons.

The magnetic moments of the $J^P=\frac{1}{2}^+$ octet baryons have been accurately measured \cite{pdg}. Our information about the $J^P=\frac{3}{2}^+$ decuplet baryons, however, is limited because of the difficulty in measuring their properties experimentally on account of their short lifetimes. Among all the decuplet baryons in hadron spectrum, $\Omega^-$ hyperon is unique as the naive SU(6) quark model describes it as a state with three strange quarks in a totally symmetric flavor-spin space.  As the strange quarks decay via the weak interaction, the $\Omega^-$ baryon is significantly more stable than
other members of $J^P=\frac{3}{2}^+$ decuplet baryons, which have at least one light quark.  The magnetic moment of $\Omega^- = -2.02 \pm 0.05 ~ \mu_N$  has been measured with high precision \cite{pdg}.  In spite of the considerable progress made over past few years to determine the magnetic moments of other decuplet baryons,  there is hardly any consensus regarding the mechanisms that can contribute to it and additional refined data is needed to bridge the gap.

The magnetic moments of $J^P=\frac{3}{2}^+$ decuplet baryons have been calculated theoretically using different approaches.
The first calculation used a SU(6) symmetric quark model (NQM) \cite{nqm}.
This work was further improved by considering the individual
contributions of the quark magnetic moments, the SU(3) symmetry
breaking effects, sea quark contributions, quark orbital momentum effects,
relativistic effects \cite{rlpm,rqm,pha,slaughter,johan,hdmagnetic,kerbikov,effmass}.
Typically, these models invoke the additivity hypothesis where a
baryon magnetic moment is given by the sum of its constituent quark
magnetic moments.
The issue regarding the magnetic moments is difficult to understand since the
magnetic moments of baryons receive contributions not only from the magnetic moments carried
by the valence quarks but also from various complicated effects,
such as relativistic and exchange current effects,
pion cloud contributions, effect of the confinement on quark masses, etc.. In the absence of any consistent way to calculate these effects simultaneously, it is very difficult to know
their relative contributions.
Recently, a number of theoretical and
computational investigations involving the magnetic moment of decuplet baryons
baryon have been carried out using the relativistic quark model (RQM) \cite{rqm}, QCD-based quark model (QCDQM) \cite{pha}, effective mass scheme (EMS) \cite{effmass}, light cone QCD sum rule (LCQSR) \cite{aliev}, QCD sum
rule  (QCDSR) \cite{zhu}, Skyrme model \cite{schwesinger}, chiral quark soliton model (CQSM)
\cite{ledwig,kim}, chiral perturbation theory ($\chi$PT)
\cite{mendieta}, lattice QCD (LQCD) \cite{boinepalli,lattice2}, bag model (BM) \cite{hong-bagmodel}, large $N_c$ \cite{large-nc}, heavy baryon
chiral perturbation theory approach (HB$\chi$PT) \cite{heavy-chi-pert},  etc.. In addition, electromagnetic properties of the baryons have been extensively studied in a chiral quark model with exchange currents ($\chi$QMEC) which are necessary for constructing gauge invariant current \cite{buchmann}.

One of the important model which finds application in the nonperturbative regime of QCD is the
chiral constituent quark model ($\chi$CQM) \cite{manohar,eichten,cheng}.
The underlying idea is based on
the possibility that chiral symmetry breaking takes place at
a distance scale much smaller than the confinement scale.
The $\chi$CQM uses the effective interaction Lagrangian
approach of the strong interactions,  where, the effective degrees of freedom are the
valence quarks and the internal Goldstone bosons (GBs),
which are coupled to the valence quarks \cite{cheng,johan,song,hd,hds}.
The $\chi$CQM with spin-spin generated configuration mixing
\cite{dgg,isgur,yaouanc,mgupta} is able to give the satisfactory explanation for
the spin and flavor distribution functions including the strangeness content of the nucleon \cite{song,hd},
weak vector and axial-vector form factors \cite{nsweak}, magnetic moments of
octet baryons, their
transitions and Coleman-Glashow sum rule \cite{hdmagnetic}, magnetic moments of octet
baryon resonances \cite{nres}, magnetic moments of $\Lambda$ resonances \cite{torres}, charge radii and quadrupole moment \cite{charge-radii}, etc..
The model is successfully extended to
predict the important role played by the small intrinsic charm content in the nucleon spin in the SU(4) $\chi$CQM  and to
calculate the magnetic moment and charge radii of charm baryons including
their radiative decays \cite{hdcharm}.
In view of the above developments in the $\chi$CQM, it becomes desirable to extend the model to calculate the magnetic moment of the $J^P=\frac{3}{2}^+$ decuplet baryons as the knowledge of magnetic moments of decuplet baryons undoubtedly provide vital clues to the spin structure and the nonperturbative aspects of QCD.

The purpose of the present paper is to formulate in detail the
magnetic moment of the $J^P=\frac{3}{2}^+$ decuplet baryons in
the  SU(3) framework of $\chi$CQM with explicit contributions
coming from the valence spin polarization, quark sea
polarization, and its orbital angular momentum. In order to
understand the implications of non valence quarks and to make our
analysis more responsive, it would also be interesting to examine the
effects of chiral symmetry breaking and SU(3) symmetry breaking parameters on the magnetic moment. Further, we would also like to study the implications of variation of the quark masses, arising due to confinement of quarks on the magnetic moments \cite{effmass,effm}.

The plan of work is as follows. To facilitate discussion, in Sec.
\ref{cccm},  chiral symmetry breaking and SU(3) symmetry breaking in the context of $\chi$CQM is revisited with an emphasis on
the importance of the sea quarks. In Sec. \ref{magmom}, we present
the essential details of the spin structure to obtain the explicit contributions coming from the valence quark polarizations, sea quark polarizations, and their orbital angular momentum for the magnetic
moments of the $J^P=\frac{3}{2}^+$ decuplet baryons.
The input parameters, numerical results and their comparison
with available data have been discussed in Sec. \ref{results}.
\section{Chiral constituent quark
model ($\chi$CQM)} \label{cccm}

The $\chi$CQM was
introduced by Weinberg and further developed by
Manohar and Georgi \cite{manohar} with the basic  idea that the set of internal
Goldstone bosons (GBs) couple directly to the valence quarks in the
interior of hadron but only at a distance scale where perturbative QCD is not applicable.

The dynamics of light quarks ($u$, $d$, and $s$) and
gluons can be described by the QCD Lagrangian
\be
{\cal{L}} = - \frac{1}{4}G_{\mu \nu}^{a} G^{\mu \nu}_{a} + i
\bar{\psi}_R \slashed{D} {\psi}_R + i \bar{\psi}_L \slashed{D}
{\psi}_L - \bar{\psi}_R M {\psi}_L - \bar{\psi}_L M {\psi}_R \,,
\label{lagrang1} \ee where $ G_{\mu \nu}^{a}$ is the gluonic gauge field strength tensor,
$D^{\mu}$ is the gauge-covariant derivative,  $M$ is the quark mass matrix and $\psi_L$  and $\psi_R$ are the left and right
handed quark fields respectively
\be \Psi_L \equiv \left(\ba{c}
u_L\\d_L\\s_L \ea \right) ~~~ {\rm and } ~~~ \Psi_R \equiv \left(\ba{c}
u_R\\d_R\\s_R \ea \right)\,. \ee
Since the mass
terms change sign  as $\psi_{R}
\to \psi_{R}$ and $\psi_{L} \to -\psi_{L}$ under the chiral transformation $(\psi \to \gamma^5 \psi)$, the Lagrangian in Eq. (\ref{lagrang1}) no
longer remains invariant. In case the mass terms in the QCD Lagrangian
are neglected, the Lagrangian will have global chiral symmetry of
the SU(3)$_L$$\times$ SU(3)$_R$ group. Since the
spectrum of hadrons in the known sector does not display parity
doublets, the chiral symmetry is believed to be spontaneously broken
around a scale of 1 GeV as
\be SU(3)_L \times
SU(3)_R \to SU(3)_{L+R} \,. \ee
As a consequence, there
exists a set of massless GBs, identified with the observed ($\pi$, $K$,
$\eta$) mesons. Within the region of QCD confinement scale
($\Lambda_{QCD} \simeq 0.1-0.3$ GeV) and the chiral symmetry breaking scale
$\Lambda_{\chi SB}$, the constituent quarks, the octet of GBs
($\pi$, K, $\eta$ mesons), and the {\it weakly} interacting gluons
are the appropriate degrees of freedom.

The effective interaction Lagrangian in this region can be expressed as
\be {\cal
L}_{{\rm int}} = \bar{\psi}(i{\slashed D} + {\slashed V})\psi +
ig_{A} \bar{\psi} {\slashed A}\gamma^{5}\psi + \cdots \,,
\label{lagrang2} \ee
where $g_{A}$ is the axial-vector coupling
constant. The gluonic degrees of
freedom can be neglected owing to small effect in the
effective quark model at low energy scale. The vector and axial-vector currents
$V_{\mu}$ and $A_{\mu}$ are defined as
\be \left( \ba{c}
  V_{\mu} \\
  A_{\mu} \\ \ea
\right)=\frac{1}{2}(\xi^{\dagger}\partial_{\mu}
\xi\pm\xi\partial_{\mu}\xi^{\dagger}),\ee
where $\xi=\mathrm{exp}( 2
i \Phi/f_{\pi})$, $f_{\pi}$ is the pseudoscalar pion decay constant
($\simeq 93$~MeV), and $\Phi$ is the field describing the dynamics
of GBs as
\bea \Phi = \left( \ba{ccc} \frac{\pi^0}{\sqrt 2} +
\beta\frac{\eta}{\sqrt 6} & \pi^+ & \alpha K^+ \\ \pi^- &
-\frac{\pi^0}{\sqrt 2} + \beta \frac{\eta}{\sqrt 6} & \alpha K^0 \\
\alpha K^- & \alpha \bar{K}^0 & -\beta \frac{2\eta}{\sqrt 6} \ea
\right)\,. \eea
Expanding $V_{\mu}$ and $A_{\mu}$ in the powers of
$\Phi/f_{\pi}$, we get
\bea
V_{\mu} &=& 0 +
O \left( (\Phi/f_{\pi})^{2} \right) \,, \\
A_{\mu} &=& \frac{i}{f_{\pi}} \partial_{\mu} \Phi + O \left(
(\Phi/f_{\pi})^{2} \right)\,.
\eea

The effective interaction Lagrangian  between
GBs and quarks from Eq. (\ref{lagrang2}) in the leading order can now be expressed as
\be {\cal
L}_{{\rm int}} = -\frac{g_{A}}{f_{\pi}} \bar{\psi} \partial_{\mu}
\Phi \gamma^{\mu} \gamma^{5} \psi \,, \label{lagrang3}
\ee
which can
be reduced to
 \be {\cal L}_{{\rm int}} \approx  i \sum_{q = u,d,s}
\frac{m_q + m_{q'}}{f_{\pi}} {\bar q'} \Phi \gamma^5 q = i
\sum_{q=u,d,s} c_8 {\bar q'} \Phi \gamma^5 q \,,
\ee
using the Dirac
equation $(i \gamma^{\mu} \partial_{\mu} - m_q)q =0$. Here,
$c_8\left( = \frac{m_q + m_{q'}}{f_{\pi}} \right)$ is the coupling
constant for octet of GBs and $m_q$ ($m_{q'}$) is the quark mass
parameter. The Lagrangian of the quark-GB interaction suppressing
all the space-time structure to the lowest order can now be
expressed as \be {\cal L}_{{\rm int}} =  c_8 {\bar \psi} \Phi \psi
\,.\ee

The QCD Lagrangian is also invariant under the axial
$U(1)$ symmetry, which would imply the existence of ninth GB. This
breaking symmetry picks the $\eta'$ as the ninth GB. The effective
Lagrangian describing interaction between quarks and a nonet of GBs,
consisting of octet and a singlet, can now be expressed as
\be {\cal
L}_{{\rm int}} = c_8 { \bar \psi} \Phi {\psi} + c_1{ \bar \psi}
\frac{\eta'}{\sqrt 3}{\psi}= c_8 {\bar \psi}\left( \Phi + \zeta
\frac{\eta'}{\sqrt 3}I \right) {\psi }=c_8 {\bar \psi} \left(\Phi'
\right) {\psi} \,, \label{lagrang4} \ee
where $\zeta=c_1/c_8$, $c_1$
is the coupling constant for the singlet GB and $I$ is the $3\times 3$
identity matrix.

The fluctuation process describing the effective Lagrangian is
\be q^{\pm} \rightarrow {\rm GB} + q^{'
\mp} \rightarrow (q \bar q^{'}) +q^{'\mp}\,, \label{basic}
\ee
where $q \bar q^{'} +q^{'}$
constitute the sea quarks \cite{cheng,johan,hd}. The GB field can be expressed in terms of the GBs and their transition probabilities as \bea
\Phi' = \left( \ba{ccc} \frac{\pi^0}{\sqrt 2}
+\beta\frac{\eta}{\sqrt 6}+\zeta\frac{\eta^{'}}{\sqrt 3} & \pi^+
  & \alpha K^+   \\
\pi^- & -\frac{\pi^0}{\sqrt 2} +\beta \frac{\eta}{\sqrt 6}
+\zeta\frac{\eta^{'}}{\sqrt 3}  &  \alpha K^o  \\
 \alpha K^-  &  \alpha \bar{K}^0  &  -\beta \frac{2\eta}{\sqrt 6}
 +\zeta\frac{\eta^{'}}{\sqrt 3} \ea \right). \eea
The transition probability of chiral
fluctuation  $u(d) \rightarrow d(u) + \pi^{+(-)}$, given in terms of the coupling constant for the octet GBs $|c_8|^2$, is defined as $a$ and is introduced by considering nondegenerate quark masses $M_s > M_{u,d}$. In terms of $a$, the probabilities of transitions of $u(d) \rightarrow s + K^{+(0)}$, $u(d,s)\rightarrow u(d,s) + \eta$, and $u(d,s) \rightarrow u(d,s) + \eta^{'}$ are given as $\alpha^2 a$, $\beta^2 a$ and $\zeta^2 a$ respectively \cite{cheng,johan}. The parameters $\alpha$ and $\beta$ are introduced by considering nondegenerate GB masses $M_{K},M_{\eta}> M_{\pi}$ and the parameter $\zeta$ is introduced by considering  $M_{\eta^{'}} > M_{K},M_{\eta}$.

\section{Magnetic moments} \label{magmom}

The magnetic moment of a given baryon in the $\chi$CQM receives
contribution from the spin of the valence quarks, spin of the sea quarks and the orbital angular motion of the sea quarks. The total magnetic moment is expressed as \be \mu(B)_{{\rm total}}= \mu(B)_{{\rm V}}+\mu(B)_{{\rm S}} + \mu(B)_{{\rm O}}\,, \label{totalmag} \ee
where $\mu(B)_{{\rm V}}$ and $\mu(B)_{{\rm S}}$ are the magnetic moment contributions of the valence quarks and the sea quarks respectively coming from their spin polarizations, whereas  $\mu(B)_{{\rm O}}$ is the magnetic moment contribution due to the rotational
motion of the two bodies constituting the sea quarks ($q^{'}$) and GB and referred to as the orbital
angular momentum contribution of the quark sea \cite{cheng}.

In terms of quark magnetic moments and spin polarizations, the
valence spin ($\mu(B)_{{\rm V}}$) , sea spin ($\mu(B)_{{\rm S}}$) , and sea orbital ($\mu(B)_{{\rm O}}$) contributions can be defined as
 \bea
\mu(B)_{{\rm V}} &=& \sum_{q=u,d,s}{\Delta q_{{\rm val}}\mu_q}\,,\label{mag-val}\\
\mu(B)_{{\rm S}} &=& \sum_{q=u,d,s} {\Delta q_{{\rm
sea}}\mu_q}\,, \label{mag-sea} \\ \mu(B)_{{\rm O}} &=& \sum_{q=u,d,s} {\Delta
q_{{\rm val}}~\mu(q_{+} \rightarrow )} \,,\label{mag-orbit}
 \eea
 where
$\mu_q= \frac{e_q}{2 M_q}$ ($q=u,d,s$) is the quark magnetic moment in the units of $\mu_N$ (nuclear magneton), $\Delta q_{{\rm val}}$
and $\Delta q_{{\rm sea}}$ are the valence and sea quark spin polarizations respectively,
$\mu(q_{+} \rightarrow )$ is the orbital moment for any chiral
fluctuation, $e_q$ and $M_q$ are the electric charge and the mass,
respectively, for the quark $q$.

The valence and sea quark spin contributions for a given baryon can be
calculated from the spin structure of the baryons. Following references \cite{cheng,johan,hd},
the quark spin polarization can be defined as
\be
\Delta q= q^{+}- q^{-},
\ee
where $q^{\pm}$ can be calculated from the spin
structure of a baryon
\be
\hat B \equiv \langle B|{\cal N}|B \rangle=\langle B|q^+q^-|B \rangle\,. \label{BNB}
\ee
Here $|B\rangle$ is the baryon wave function  and ${\cal N}=q^+q^-$ is the number
operator measuring the sum of the quark  numbers with spin up or down, for example,
\be
q^+q^-=\sum_{q=u,d,s} (n_{q^{+}}q^{+} + n_{q^{-}}q^{-})=n_{u^{+}}u^{+} + n_{u^{-}}u^{-} + n_{d^{+}}d^{+} + n_{d^{-}}d^{-} +
n_{s^{+}}s^{+} + n_{s^{-}}s^{-}\,, \label{number}
\ee
with the coefficients of the $q^{\pm}$ giving the number of
$q^{\pm}$ quarks.

The valence quarks spin polarizations ($\Delta
q_{{\rm val}}= n_{q^{+}}- n_{q^{-}}$) for a given baryon can be
calculated using the SU(6) spin-flavor wave functions $|B\rangle$ in
Eq. (\ref{BNB}).

The quark sea spin polarizations ($\Delta q_{{\rm sea}}$) coming from the fluctuation process
in Eq. (\ref{basic}) can be calculated  by substituting for every
valence quark
\be
q^{\pm} \to \sum P_q q^{\pm} + |\psi(q^{\pm})|^2, \label{sea-q}
\ee
where the transition probability of the emission of a GB from
any of the $q$ quark ($\sum P_q$)  and the transition
probability of the $q^{\pm}$ quark ($|\psi(q^{\pm})|^2$) can be calculated from the Lagrangian. They are expressed as
\[ \sum P_u= -a\left( \frac{9+\beta^2+2 \zeta^2}{6} +\alpha^2\right)~~~
{\rm and}~~~
|\psi(u^{\pm})|^2=\frac{a}{6}(3+\beta^2+2 \zeta^2)u^{\mp}+
a d^{\mp}+a \alpha^2 s^{\mp}\,,      \]

\[ \sum P_d= -a\left( \frac{9+\beta^2+2 \zeta^2}{6} +\alpha^2\right)~~~
{\rm and}~~~
|\psi(d^{\pm})|^2=a u^{\mp}+
\frac{a}{6}(3+\beta^2+2 \zeta^2)d^{\mp}+ a \alpha^2 s^{\mp}\,,
                                               \]

\[ \sum P_s= -a\left( \frac{2 \beta^2+\zeta^2}{3}+2 \alpha^2\right)~~~
{\rm and}~~~
|\psi(s^{\pm})|^2=   a \alpha^2 u^{\mp}+
a \alpha^2 d^{\mp}+\frac{a}{3}(2 \beta^2+\zeta^2)s^{\mp}\,.
 \]
The contribution of the
angular momentum of the sea quarks to the magnetic moment of a given
quark is
\be
\mu (q^{+} \rightarrow {q}^{'-}) =\frac{e_{q^{'}}}{2M_q}
\langle l_q \rangle +
\frac{{e}_{q}-{e}_{q^{'}}}{2 {M}_{{\rm GB}}}\langle {l}_{{\rm GB}} \rangle\,,
\ee
where
\be
\langle l_q \rangle=\frac{{M}_{{\rm GB}}}{M_q+{M}_{{\rm GB}}} ~{\rm and}
~\langle l_{{\rm GB}} \rangle=\frac{M_q}{M_q+{M}_{{\rm GB}}}\,,
\ee
$\langle l_q, l_{{\rm GB}} \rangle$ and ($M_q$, ${M}_{{\rm GB}}$) are the
orbital angular momenta and masses of quark and GB respectively.
The orbital moment of each process is then multiplied by the probability
for such a process to take place to yield the magnetic moment due to
all the transitions starting with a given valence quark, for example
\bea
 [ \mu (u^{\pm}(d^{\pm}) \rightarrow )] &=&  \pm a
[\mu \left(u^{+}(d^{+}) \rightarrow d^- (u^-)\right) +
\alpha^2 \mu \left(u^+(d^+) \rightarrow s^-\right) \nonumber \\
 &+& \left.\left(\frac{1}{2} +\frac{1}{6} \beta^2+ \frac{1}{3} \zeta^2\right)
\mu \left(u^{+}(d^{+}) \rightarrow u^- (d^-)\right)\right], \label{mud}
\eea
\be
[\mu (s^{\pm} \rightarrow )] =  \pm a
\left[\alpha^2 \mu \left(s^{+} \rightarrow u^-\right) +
\alpha^2 \mu \left(s^+ \rightarrow d^-\right) +
\left(\frac{2}{3} \beta^2+ \frac{1}{3} \zeta^2\right)
\mu \left(s^{+} \rightarrow s^- \right)\right]. \label{mus}
\ee
The above equations can easily be
generalized by including the coupling breaking and mass
breaking terms, for example, in terms of the coupling breaking
parameters $a$, $\alpha$, $\beta$ and $\zeta$  as well as the masses
of GBs $M_{\pi}$, $M_{K}$
and $M_{\eta}$. The  orbital moments of $u$, $d$ and $s$ quarks
respectively are
\be [\mu(u^+ \rightarrow)] = a \left [\frac{3 M^2_{u}}{2
{M}_{\pi}(M_u+ {M}_{\pi})}- \frac{\alpha^2(M^2_{K}- 3 M^2_{u})}{2
{M}_{K}(M_u+ {M}_{K})} + \frac{\beta^2 M_{\eta}}{6(M_u+
{M}_{\eta})}+ \frac{\zeta^2 M_{\eta'}}{3(M_u+ {M}_{\eta'})}  \right]
{\mu}_u \,, \label{orbitu} \ee
\be [\mu(d^+ \rightarrow)] = -2 a
\left [\frac{3( M^2_{\pi}-2 M^2_{d})}{4 {M}_{\pi}(M_d+ {M}_{\pi})}-
\frac{\alpha^2 M_{K}}{2(M_d+ {M}_{K})}  - \frac{\beta^2
M_{\eta}}{12(M_d+ {M}_{\eta})}- \frac{\zeta^2 M_{\eta'}}{6(M_d+
{M}_{\eta'})} \right ] {\mu}_d \,, \label{orbitd} \ee
\be [\mu(s^+
\rightarrow)] = -2 a \left[ \frac{\alpha^2 (M^2_{K}-3
M^2_s)}{2{M}_{K}(M_s+ {M}_{K})}  - \frac{\beta^2 M_{\eta}}{3(M_s+
{M}_{\eta})}  - \frac{\zeta^2 M_{\eta'}}{6(M_s+ {M}_{\eta'})} \right
]{\mu}_s\,. \label{orbits} \ee

The orbital contribution to the magnetic moment of the baryon of the type
$B(q_1 q_2 q_3)$ is given as
\be
\mu(B)_{{\rm O}} =  \Delta {q_1}_{{\rm val}} [\mu ({q_1}^+ \rightarrow)] +
\Delta {q_2}_{{\rm val}}[\mu ({q_2}^+ \rightarrow)]
+\Delta {q_3}_{{\rm val}} [\mu ({q_3}^+ \rightarrow)]\,. \label{de orbit}
\ee

We now discuss in detail the valence quark spin, sea quark spin and sea quark orbital contributions to
the magnetic moment of  $J^P=\frac{3}{2}^+$ decuplet baryons ($B^*$).
To calculate $\mu(B^*)_{{\rm V}}$, we need to calculate the valence
spin polarizations $\Delta q^{B^{*}}_{{\rm val}}$ from the valence spin structure in a totally symmetric
flavor-spin-space from Eq. (\ref{BNB})
\be
\widehat{B^*} \equiv \langle B^*(J^P=\frac{3}{2}^+)|{\cal N}|B^*(J^P=\frac{3}{2}^+) \rangle \,. \label{valo}
 \ee
The spin structure for the $J^P=\frac{3}{2}^+$ decuplet baryons is expressed as
\bea
\Delta^{++} &= & 3 u^+ +0 u^-+0 d^+ + 0 d^- + 0 s^+ + 0 s^-,  \nonumber \\
\Delta^{+}&= & 2 u^+ +0 u^-+1 d^+ + 0 d^- + 0 s^+ + 0 s^-,  \nonumber \\
\Delta^{o}&= & 1 u^+ +0 u^-+2 d^+ + 0 d^- + 0 s^+ + 0 s^-,  \nonumber \\
\Delta^{-}&= & 0 u^+ +0 u^-+3 d^+ + 0 d^- + 0 s^+ + 0 s^-,  \nonumber \\
\Sigma^{*^+} &=& 2 u^+ +0 u^-+0 d^+ + 0 d^- + 1 s^+ + 0 s^-,  \nonumber \\
\Sigma^{*^o}&=& 1 u^+ +0 u^-+1 d^+ + 0 d^- + 1 s^+ + 0 s^-,  \nonumber \\
\Sigma^{*^-} &=& 0 u^+ +0 u^-+2 d^+ + 0 d^- + 1 s^+ + 0 s^-,  \nonumber \\
\Xi^{*^o}  &=& 1 u^+ +0 u^-+0 d^+ + 0 d^- + 2 s^+ + 0 s^-,  \nonumber \\
\Xi^{*^-} &=& 0 u^+ +0 u^-+1 d^+ + 0 d^- + 2 s^+ + 0 s^-,  \nonumber \\
\Omega^{-} &=& 0 u^+ +0 u^-+0 d^+ + 0 d^- + 3 s^+ + 0 s^-. \label{spin-structure-hyperons}
\eea
The resulting spin polarizations give the
valence contribution to the magnetic moment obtained by substituting these in Eq. (\ref{mag-val}). The results have been presented in Table \ref{tabdecval}.
\begin{table}
\begin{center}
{\renewcommand{\arraystretch}{1.4}
\tabcolsep 5mm
\begin{tabular}{|cc|}     \hline \hline
$B^*(J^P=\frac{3}{2}^+)$& $\mu(B^*)_{{\rm V}}$ \\ \hline \hline

$\Delta^{++}(uuu)$ & $3 \mu_{u}$  \\
$\Delta^{+}(uud)$ & $2 \mu_{u}+ \mu_{d}$  \\
$\Delta^{o}(udd)$ & $ \mu_{u}+ 2\mu_{d}$ \\
$\Delta^{-}(ddd)$ & $3 \mu_{d}$  \\
$\Sigma^{*^+}(uus)$ &$2 \mu_{u}+ \mu_{s}$  \\
$\Sigma^{*^o}(uds)$ & $\mu_{u}+ \mu_{d}+\mu_{s}$ \\
$\Sigma^{*^-}(dds)$ & $ 2\mu_{d}+\mu_{s}$ \\
$\Xi^{*^o}(uss)$  & $\mu_{u}+ 2\mu_{s}$ \\
$\Xi^{*^-}(dss)$  & $\mu_{d}+2\mu_{s}$ \\
$\Omega^{-}(sss)$ & $3\mu_{s}$\\ \hline \hline
\end{tabular}
\caption{Valence contribution to the magnetic moment $\mu(B^*)_{{\rm V}}$ for the $J^P=\frac{3}{2}^+$ hyperons. }
\label{tabdecval}}
\end{center}
\end{table}

The sea quark polarizations which contribute to the magnetic moment of $J^P=\frac{3}{2}^+$ decuplet baryons can be
calculated by substituting Eq. (\ref{sea-q}) for every valence quark in Eq. (\ref{spin-structure-hyperons}). Consequently, the magnetic moment contributions of the sea quarks $\mu(B^*)_{{\rm S}}$ can be calculated from Eq. (\ref{mag-sea}) and the results have been presented in  Table \ref{decsea}.

\begin{table}
\begin{center}
\begin{tabular}{|cc|}     \hline \hline
$B^*(J^P=\frac{3}{2}^+)$ & $\mu(B^*)_{{\rm S}}$ \\ \hline \hline
$\Delta^{++}(uuu)$ & $- a\left[ \left( 6 + 3 \alpha ^2 +\beta^2+2 \zeta^2 \right) \mu_u +3 \mu_d +3 \alpha^2 \mu_s \right]$ \\
$\Delta^{+}(uud)$ & $- a\left[\left( 5+2 \alpha ^2 + \frac{2}{3} \beta^2+\frac{4}{3} \zeta^2 \right) \mu_u+ \left( 4 + \alpha ^2 + \frac{1}{3} \beta^2+\frac{2}{3}\zeta^2 \right) \mu_d+3 \alpha^2 \mu_s \right]$ \\
$\Delta^{o}(udd)$ & $- a\left[\left( 4 + \alpha ^2 + \frac{1}{3} \beta^2+\frac{2}{3}\zeta^2\right) \mu_u+ \left( 5 + 2 \alpha ^2 + \frac{2}{3} \beta^2+\frac{4}{3} \zeta^2\right) \mu_d+3 \alpha^2 \mu_s\right]$ \\
$\Delta^{-}(ddd)$ & $-a \left[3 \mu_u +\left( 6 + 3 \alpha ^2 +\beta^2+2 \zeta^2\right)\mu_d+3 \alpha^2 \mu_s\right]$ \\
$\Sigma^{*^+}(uus)$ & $- a\left[ \left( 4+3 \alpha^2 + \frac{2}{3} \beta^2+\frac{4}{3}\zeta^2\right) \mu_u
 +\left( \alpha ^2 + 2\right)\mu_d+2 \left( 2 \alpha ^2 + \frac{2}{3} \beta^2+\frac{1}{3} \zeta^2 \right) \mu_s \right]$ \\
$\Sigma^{*^o}(uds)$ & $- a\left[ \left( 3 + 2\alpha ^2 +\frac{1}{3} \beta^2+\frac{2}{3} \zeta^2 \right) \mu_u +\left( 3 + 2\alpha ^2 +\frac{1}{3} \beta^2+\frac{2}{3} \zeta^2\right)\mu_d + 2\left( 2\alpha ^2 +\frac{2}{3} \beta^2+\frac{1}{3} \zeta^2\right)\mu_s\right]$ \\
$\Sigma^{*^-}(dds)$ & $- a\left[ \left( \alpha ^2 + 2 \right) \mu_u+\left( 4 + 3 \alpha ^2 + \frac{2}{3} \beta^2+\frac{4}{3} \zeta^2 \right)\mu_d+
2\left( 2 \alpha ^2 + \frac{2}{3} \beta^2+\frac{1}{3} \zeta^2\right)\mu_s \right]$ \\
$\Xi^{*^o}(uss)$  &  $-a \left[ \left(2+3 \alpha^2+\frac{1}{3}\beta^2+\frac{2}{3}\zeta^2\right)\mu_u+\left(2 \alpha^2 +1\right)\mu_d+ \left(5 \alpha^2 + \frac{8}{3} \beta^2+\frac{4}{3} \zeta^2\right)\mu_s \right]$ \\
$\Xi^{*^-}(dss)$  &   $-a\left[ \left(2 \alpha^2 +1\right)\mu_u+ \left(2+3 \alpha^2+\frac{1}{3}\beta^2+\frac{2}{3} \zeta^2\right)\mu_d+\left(5 \alpha^2 + \frac{8}{3} \beta^2+\frac{4}{3} \zeta^2\right)\mu_s\right]$ \\
$\Omega^{-}(sss)$ &  $-3a\left[ \alpha^2 \mu_u+ \alpha^2 \mu_d +2\left(\alpha^2 + \frac{2}{3} \beta^2+\frac{1}{3} \zeta^2 \right)\mu_s \right]$ \\  \hline \hline
\end{tabular}
\caption{Sea contribution to the magnetic moment $\mu(B^*)_{{\rm S}}$ for the $J^P=\frac{3}{2}^+$ decuplet baryons. }
\label{decsea}\end{center}
\end{table}

The orbital contribution to the total magnetic moment of the $J^P=\frac{3}{2}^+$ decuplet baryons, as given by Eq. (\ref{de orbit}),  is expressed as
\bea
\mu(\Delta^{++})_{{\rm O}} &=& 3 \mu(u^+ \rightarrow),  \nonumber \\
\mu(\Delta^{+})_{{\rm O}} &=& 2 \mu(u^+ \rightarrow) + \mu(d^+ \rightarrow),  \nonumber \\
\mu(\Delta^{o})_{{\rm O}} &=&  \mu(u^+ \rightarrow) + 2 \mu(d^+ \rightarrow),  \nonumber \\
\mu(\Delta^{-})_{{\rm O}} &=& 3 \mu(d^+ \rightarrow),  \nonumber \\
\mu(\Sigma^{*^+})_{{\rm O}} &=& 2 \mu(u^+ \rightarrow) + \mu(s^+ \rightarrow),  \nonumber \\
\mu(\Sigma^{*^o})_{{\rm O}}&=& \mu(u^+ \rightarrow)+\mu(d^+ \rightarrow)+\mu(s^+ \rightarrow),  \nonumber \\
\mu(\Sigma^{*^-})_{{\rm O}} &=& 2\mu(d^+ \rightarrow)+\mu(s^+ \rightarrow),  \nonumber \\
\mu(\Xi^{*^o})_{{\rm O}}  &=& \mu(u^+ \rightarrow)+2\mu(s^+ \rightarrow),  \nonumber \\
\mu(\Xi^{*^-})_{{\rm O}} &=& \mu(d^+ \rightarrow)+2\mu(s^+ \rightarrow),  \nonumber \\
\mu(\Omega^{-})_{{\rm O}} &=& 3\mu(s^+ \rightarrow). \label{orbital-structure-hyperons}\eea
Using Eqs. (\ref{orbitu}), (\ref{orbitd})  and (\ref{orbits}), the orbital contribution to the total magnetic moment of the $J^P=\frac{3}{2}^+$ decuplet baryons can be calculated.

\section{Effective Quark  Masses}

The basic assumptions of $\chi$CQM suggest that  the constituent
quarks are supposed to have only Dirac magnetic moments governed by
the respective quark masses. Since we do not have any definite guidelines for the constituent quark
masses, we can use the most widely accepted values of quark masses in hadron spectroscopy. In order to study the effect of quark confinement on the magnetic moments, the effective quark masses can be considered \cite{effmass,effm}. In conformity with  additivity assumption,
the simplest way to incorporate this adjustment
is to first express $M_q$ in the magnetic moment
operator in terms of $M_B$, the mass of the baryon obtained
additively from the quark masses, which then
is replaced  by $M_B+\Delta M$,
$\Delta M$
being the mass difference between the experimental value and $M_B$.
This leads to the following modification in the quark
magnetic moments
 \bea
\mu_u^{\rm eff} &=& 2[1-(\Delta M/M_B)] {\mu}_N, \nonumber\\
\mu_d ^{\rm eff}&=& -[1-(\Delta M/M_B)] {\mu}_N,\nonumber \\
\mu_s^{\rm eff} &=& -M_u/M_s [1-(\Delta M/M_B)]{\mu}_N.
\eea
In addition to this, various sum rules derived from the spin-spin
interactions for different baryons {\cite{{isgur},{yaouanc},{mgupta}}}
can be used to fix $M_s$, for example,
$(\Sigma^*-\Sigma)/(\Delta-N)=M_u/M_s$ and
$(\Xi^*-\Xi)/(\Delta-N)=M_u/M_s$.
The baryon magnetic moments calculated after incorporating this effect
would be referred to as baryon magnetic moments with effective quark masses.

\section{Results and Discussion}
\label{results}

The explicit numerical values of the valence, sea, and orbital contributions to the magnetic moment of $J^P=\frac{3}{2}^+$ decuplet baryons in $\chi$CQM can be calculated. The magnetic moment calculations in $\chi$CQM with SU(3) broken symmetry involve the symmetry breaking parameters $a$, $a\alpha^2$, $a \beta^2$, $a \zeta^2$,
representing, respectively, the probabilities of fluctuations of a constituent quark into pions, $K$, $\eta$, $\eta^{'}$. These parameters provide the basis to understand the extent to which the sea quarks and their orbital angular momentum contribute to the structure of the baryon. The hierarchy for the probabilities, which scale as $\frac{1}{M_q^2}$, can be obtained as
\be a> a \alpha^2 \geq a \beta^2> a \zeta^2.
\ee
Since the parameters cannot be fixed independently, therefore, we have carried out a broader analysis
to find the ranges of the $\chi$CQM parameters from experimentally well known quantities pertaining to the spin polarization functions and quark distribution
functions.
The range of the coupling breaking parameter $a$ can be easily
found by considering the expressions of the quark spin polarization
functions by giving the full variation
of parameters $\alpha$, $\beta$ and $\zeta$ \cite{hd}. We obtain $0.10 \lesssim a \lesssim 0.14$.
The range of the parameter $\zeta$ can
be found from the latest experimental
measurement  of $\bar u/\bar d$ \cite{e866} which involves only $\beta$ and $\zeta$. Using the possible range
of $\beta$, i.e. $0<\beta<1$ one finds $-0.70
\lesssim \zeta \lesssim -0.10$. The range of $\beta$ can be found by using the
$\bar u-\bar d$ asymmetry representing the violation of Gottfried
sum rule \cite{GSR} and expressed as $\bar u-\bar
d=\frac{a}{3}(2 \zeta+\beta-3)$. Using the above
found ranges of $a$ and $\zeta$ as well as the latest measurement
of $\bar u-\bar d$ asymmetry \cite{e866}, $\beta$ falls in the range $0.2\lesssim \beta \lesssim 0.7$. Similarly, the
range of $\alpha$ can be found by considering the flavor
non-singlet component $\Delta_3$ (=$\Delta u-\Delta d$) and it
comes out to be $0.2 \lesssim \alpha \lesssim 0.5$.

In
addition to the parameters of $\chi$CQM as discussed above, the
orbital angular momentum contributions are characterized by the
quark and GB masses. For evaluating their contribution, we have used
their on shell mass values in accordance with several other
similar calculations.
In particular, for the constituent quark masses $u$, $d$, and $s$ we
have used their widely accepted values in hadron spectroscopy
$M_u=M_d=330$ MeV, $M_s=510$ MeV  \cite{cheng,isgur,mu1}.
In addition, the effect of the confinement on quark masses can be added by taking the effective quark magnetic moments $\mu_q^{\rm eff}$ as discussed in the previous section.

This model has already been applied to calculate the magnetic moments
of the octet baryons \cite{hdmagnetic}
where experimental data is available for all the cases. It
is interesting to observe that our results for the magnetic moments
of $p$, $\Sigma^{+}$, $\Xi^{0}$, and $\Lambda$ give a perfect fit to
the experimental values \cite{pdg} whereas for all other octet
baryons our predictions are within 10\% of the observed values.
Besides this, we have also been able to get an excellent fit to
violation of Coleman-Glashow sum rule \cite{cgsr}.
The fit becomes all the more impressive when it is
realized that none of the magnetic moments are used as inputs and
the violation of Coleman-Glashow sum rule can be described without resorting to additional
parameters. In all the cases, the contribution of the quark sea and its orbital angular
momentum is quite significant when compared with the valence
contribution.

Using the inputs discussed above and performing a full scan of the $\chi$CQM parameters at $1 \sigma$ CL, in Table  \ref{hyperons} we have
presented the $\chi$CQM results of the $J^P=\frac{3}{2}^+$ decuplet baryons magnetic moments. In order to study closely the role of confinement on quark masses on the magnetic moments, in the table we have also presented the
$\chi$CQM results   with the effective quark magnetic moments $\mu_q^{\rm eff}$.
One can immediately see that
the total contribution to the magnetic moment is coming from different sources with similar and
opposite signs, for example, the orbital is contributing with the same
sign as the valence part, whereas the sea is contributing with opposite
sign. For example, in the case of $\mu(\Delta^-)$ and $\mu(\Sigma^-)$,
because the orbital part dominates over the sea quark
polarization,  the magnetic moments are higher as compared to just the valence contributions. On the other hand, in the case of  $\mu(\Delta^+)$ and $\mu(\Sigma^+)$, the
sea quark polarization dominates over the orbital part as a consequence of
which the magnetic moment contribution is more or less the same as
that of the the valence contribution.
In general, one can find that whenever there is an excess of $d$ quarks
the orbital part dominates, whereas when we have an excess of $u$ quarks,
the sea quark polarization dominates.
A measurement of these magnetic moments, therefore, would have
important implications for the $\chi$CQM.
The sea and orbital contributions, in a very interesting manner, add on to the valence contributions leading to better agreement
with data. This clearly suggests that the
sea quarks and their orbital contributions could perhaps provide the dominant dynamics of the
constituents in the nonperturbative regime of QCD on
which further corrections could be evaluated.

Since the experimental data is available for $\Delta^{++}$, $\Delta^{+}$ and
$\Omega^-$, it would be interesting to compare these with the $\chi$CQM results as well as with the results of other models. From
the table, it is clearly evident that a very
good agreement pertaining to the case of  $\mu(\Delta^{+})$  is obtained. The result for $\mu(\Delta^{++})$ also lies very well within the available range. In both the cases, the sea and orbital contributions are quite significant. They cancel in the right direction and with the right magnitude to give the total magnetic moments. The magnetic moment of $\Omega^{-}$ agrees with the experimentally observed value $-2.02 \pm 0.05$ \cite{pdg}.
Since there is an excess of strange quarks in the valence structure of $\Omega^{-}$, the contribution of the quark sea
and its orbital angular momentum is almost negligible as compared to
the valence contribution. This is due to the fact that the strange contribution to the magnetic moment is almost an
order of magnitude smaller than the up and down quarks thus leading to a very small contribution from the heavy quarks when compared
with the contribution coming from the light quarks. This becomes more clear when we study the implications of SU(3) symmetry breaking. When we carry out the
calculations in SU(3) symmetry with $\alpha=\beta=-\zeta=1$, we obtain
$\mu(\Omega^-)_{{\rm V}}=-1.94$, $\mu(\Omega^-)_{{\rm S}}=0.60$ and $\mu(\Omega^-)_{{\rm O}}=-0.20$ giving
$\mu(\Omega^-)_{{\rm total}}= -1.54$ which is even worse than just the valence contribution. The result with SU(3) symmetry is also clearly in disagreement with the experimentally observed value $-2.02 \pm 0.05$ \cite{pdg}.
Thus, SU(3) symmetry breaking plays an important role in obtaining the fit. 
Another interesting observation for the case of $\Omega^{-}$ is that the central value of $\mu(\Omega^-)$ is not equal to the sum of central values of $\mu(B^*)_{{\rm V}}$, $\mu(B^*)_{{\rm S}}$ and $\mu(B^*)_{{\rm O}}$ for the $\chi$CQM results with the effective quark magnetic moments. This is primarily because of the valence structure of $\Omega^{-}$ which has three strange quarks. The orbital contribution of $\Omega^{-}$ which comes purely from the orbital moment of $s$ quarks plays a very crucial role here. On having a closer look at the expression of the orbital moment of $s$ quarks from Eq. (\ref{orbits}), we find that, unlike the orbital moments of $u$ and $d$ quarks (Eqs. (\ref{orbitu}) and (\ref{orbitd})) where the major numerical contribution comes from the mass terms without $\alpha^2$, $\beta^2$ and $\zeta^2$, the orbital moment of $s$ quark has no such term. This yields a rather large value of error (approx $50\%$) in the result of the orbital contribution given in Eq. (\ref{orbital-structure-hyperons}). This is not observed in any other baryon.
For the sake of comparison, we have presented the results of other available phenomenological and theoretical models in Table \ref{data}.
Our model predictions for all the decuplet baryons are more or less in agreement with the predictions of other models existing in literature. In the case of $\Delta^{++}$, $\Delta^{+}$ and
$\Omega^-$ where data is available, our results are even better than most of the predictions of other models.

\begin{table}
\hspace {2cm}
\tabcolsep 0.5mm
{\renewcommand{\arraystretch}{1.4}
\hbox{
\begin{sideways}
\begin{tabular}{|c|c|c|c|c|c|c|c|c|c|}      \hline \hline
 &   &
\multicolumn{4}{c|} {$\chi$CQM} &
\multicolumn{4}{c|} {$\chi$CQM with $\mu_q^{\rm eff}$}\\
\cline{3-6} \cline {7-10}
Baryons &  Data  & $\mu(B^*)_{{\rm V}}$  & $\mu(B^*)_{{\rm S}}$ &$\mu(B^*)_{{\rm O}}$ &  $\mu(B^*)$ &  $\mu(B^*)_{{\rm V}}$  & $\mu(B^*)_{{\rm S}}$ & $\mu(B^*)_{{\rm O}}$ &  $\mu(B^*)$ \\
\hline \hline

$\mu(\Delta^{++})$ & $3.7 <\mu(\Delta^{++}) <7.5$ \cite{pdg}& $6$ & $-1.32 \pm 0.15$  & 1.07$\pm$0.12 & 5.82$\pm$0.08  & 6.06 &$-1.31\pm0.15$  & 1.08$\pm$0.12     & 5.82$\pm$0.08  \\
$\mu(\Delta^{+})$ &$2.7^{+1.0}_{-1.3}\pm1.5\pm3$ \cite{prl-deltaplus}& $3$ & $-0.83\pm0.09$  & 0.42$\pm$0.06     & 2.63$\pm$0.06 & 3.03  &$-0.82\pm0.09$  & 0.42$\pm$0.06  & 2.63$\pm$0.06   \\

$\mu(\Delta^{o})$ &-- & $0$ & $-0.33\pm0.04$  &$-0.23\pm0.03$  & $-0.56\pm0.09$  & 0  &$-0.33\pm0.05$  & $-0.23\pm0.03$    & $-0.55\pm0.09$ \\

$\mu(\Delta^{-})$ &-- & $-3$ & $0.16\pm0.05$  & $-0.85\pm0.11$  & $-3.75\pm0.08$  & $-3.03$  &$0.16\pm0.05$  & $-0.87\pm0.09$     & $-3.75\pm0.08$ \\

$\mu(\Sigma^{*+})$ &-- & $3.37$ & $-0.85\pm0.08$  & 0.68$\pm$0.06 & 3.25$\pm$0.05 & 3.24 &$-0.81\pm0.06$  & 0.658$\pm$0.07     & 3.09$\pm$0.03 \\

$\mu(\Sigma^{*o})$ &  -- &  0.37 & $-0.35\pm0.03$  & 0.018$\pm$0.004 &
0.05$\pm$0.03 & 0.33 & $-0.33\pm0.03$  & 0.018$\pm$0.004    &
0.018$\pm$0.03 \\

$\mu(\Sigma^{*-})$ & --  &  $-2.63$ & 0.14$\pm$0.04  & $-0.62\pm0.07$ & $-3.14\pm0.06$ & $-2.58$ & 0.14$\pm$0.04  & $-0.62\pm0.07$
   & $-3.07\pm0.06$ \\

$\mu(\Xi^{*o})$    & -- &  0.74 & $-0.36\pm0.04$  & 0.26$\pm$0.04 & 0.65$\pm$0.03 & 0.52 & $-0.32\pm0.04$ & 0.26$\pm$0.04    & 0.46$\pm$0.03 \\

$\mu(\Xi^{*-})$    &  --&  $-2.26$ & 0.13$\pm$0.04  &$-0.38\pm0.06$ & $-2.53\pm0.06$ & $-2.30$ & $0.13\pm0.05$  & $-0.38\pm0.06$    &
 $-2.55\pm0.05$\\

$\mu(\Omega^-)$ & $-2.02\pm$0.05 \cite{pdg} &  $-1.89$ & 0.12$\pm$0.04  &$-0.14\pm0.07$ & $-1.92\pm0.07$ &$-1.98$ & 0.13$\pm$0.05  & $-0.14\pm0.07$    & $-2.08\pm0.08$ \\   \hline \hline

\end{tabular}
\end{sideways}
\hspace {0.6cm}
\rotcaption{Magnetic moments in units of $\mu_N$ for the $J^P=\frac{3}{2}^+$ decuplet baryons.} \label{hyperons}}}
\end{table}

\begin{table}
\tabcolsep 0.5mm
\hbox{
\begin{sideways}{\footnotesize
\begin{tabular}{|l|c|c|c|c|c|c|c|c|c|c|} \hline \hline Other models  & $\mu(\Delta^{++})$ & $\mu(\Delta^{+})$ & $\mu(\Delta^{o})$ & $\mu(\Delta^{-})$ & $\mu(\Sigma^{*+})$ & $\mu(\Sigma^{*o})$ & $\mu(\Sigma^{*-})$ & $\mu(\Xi^{*o})$ & $ \mu(\Xi^{*-})$&$\mu(\Omega^-)$ \\  \hline \hline
NQM \cite{nqm} &5.56& 2.73& $-0.09$& $-2.92$& 3.09& $0.27$& $-2.56$&$-0.63$& $-2.2$& $-1.84$ \\
RQM \cite{rqm} &4.76& 2.38& 0.00& $-2.38$& 1.82& $-0.27$& $-2.36$&$-0.60$& $-2.41$& $-2.48$ \\
QCDQM \cite{pha} &5.689& 2.778& $-0.134$& $-3.045$& 2.933& 0.137& $-2.659$& 0.424& $-2.307$& $-1.970$ \\
EMS \cite{slaughter} &3.67$\pm$.07&1.83$\pm$0.04&0$\pm$0&$-1.83\pm$0.04&1.89$\pm$0.04&0$\pm$0&$-1.89\pm$0.04&0$\pm$0&$-1.95\pm$0.05&$-2.02\pm$0.05 \\
$\chi$QM \cite{johan} &5.30& 2.58& $-0.13$& $-2.85$& 2.88& $0.17$& $-2.55$& 0.47& $-2.25$& $-1.95$ \\
EMS \cite{effmass} &4.56& 2.28& $0$& $-2.28$& 2.56& 0.23& $-2.10$& 0.48& $-1.90$& $-1.67$ \\
LCQSR \cite{aliev} &4.4$\pm$0.8&2.2$\pm$0.4&0.0&$-2.2\pm$0.4&2.7$\pm$0.6&0.20$\pm$0.05&$-2.28\pm$0.5&0.40$\pm$0.08&$-2.0\pm$0.4&$-1.65\pm$0.35 \\
QCDSR \cite{zhu} &$ 4.39\pm1.00$ & $2.19\pm0.50$ & 0.00 & $-2.19\pm0.50$ & $2.13\pm0.82$ & $0.32\pm0.15$ & $-1.66\pm0.73$ & $-0.69\pm0.29$& $-1.51\pm0.52$& $-1.49\pm0.45$ \\
CQSM \cite{ledwig} &4.85 & 2.35 & $-0.14$ & $-2.63$ &  2.47& $-0.02$ & $-2.52$ &  0.09& $-2.40$ & $-2.29$ \\
CQSM \cite{kim} &4.73& 2.19& $-0.35$& $-2.90$& 2.52& $-0.08$& $-2.69$& 0.19& $-2.48$& $-2.27$ \\
$\chi$PT \cite{mendieta} &5.390& 2.383& $-0.625$& $-3.632$& 2.519& $-0.303$& $-3.126$& 0.149& $-2.596$& $-2.042$ \\
LQCD \cite{boinepalli} &$4.91\pm0.61$& $2.46\pm0.31$ & 0.00& $-2.46\pm0.31$& $2.55\pm0.26$& $0.27\pm0.05$& $-2.02\pm0.18$& $0.46\pm0.07$& $-1.68\pm0.12$& $-1.40\pm0.10$ \\
LQCD \cite{lattice2} &5.24(18)& 0.97(65)& $-0.035(2)$& $-2.98(19)$& 1.27(6)& $0.33(5)$& $-1.88(4)$&$0.16(4)$& $-0.62(1)$& $--$ \\
CBM \cite{hong-bagmodel} &4.52& 2.12& $-0.29$& $-2.69$& 2.63& 0.08& $-2.48$& 0.44& $-2.27$& $-2.06$ \\
Large $N_c$ \cite{large-nc} &5.9(4)& 2.9(2)& $--$& $-2.9(2)$& 3.3(2)& $0.3(1)$& $-2.8(3)$&$0.65(20)$& $-2.30(15)$& $-1.94$ \\
HB$\chi$PT \cite{heavy-chi-pert} &4.0(4)& 2.1(2)& $-0.17(4)$& $-2.25(19)$& 2.0(2)& $-0.07(2)$& $-2.2(2)$&$0.10(4)$& $-2.0(2)$& $-1.94$ \\
$\chi$QMEC \cite{buchmann} &6.93& 3.47& $0.00$& $-3.47$& 4.12& $0.53$& $-3.06$& 1.10& $-2.61$& $-2.13$ \\
 \hline
This work: $\chi$CQM& 5.82$\pm$0.08 &2.63$\pm$0.06 &$-0.56\pm$0.09 & $-3.75\pm$0.08 & 3.25$\pm$0.05 & 0.05$\pm$0.03  & $-3.14\pm$0.06& 0.65$\pm$0.03&$-2.53\pm$0.06&$-1.92\pm$0.07\\

{$\chi$CQM with $\mu_q^{\rm eff}$}& 5.82$\pm$0.08&2.63$\pm$0.06&$-0.55\pm$0.09&$-3.75\pm$0.08&3.09$\pm$0.03&0.018$\pm$0.03&$-3.07\pm$0.06&0.46$\pm$0.03&$-2.55\pm$0.05&$-2.09\pm$0.08 \\ \hline \hline \end{tabular}}

\end{sideways}
\hspace {0.6cm}
\rotcaption{Phenomenological results of some other theoretical
approaches for magnetic moments of the $J^P=\frac{3}{2}^+$ decuplet baryons.} \label{data}}
\end{table}

To summarize, in a very interesting manner, the  chiral constituent quark model ($\chi$CQM) is able to phenomenologically estimate the the orbital and sea
contributions  to the magnetic moment of the $J^P=\frac{3}{2}^+$ decuplet baryons. These results along with the valence contributions lead
to a better agreement with data. The magnetic moments have implications for chiral symmetry
breaking and SU(3) symmetry breaking and provide a basis to understand the significance of sea quarks in the baryon structure. The results also suggest
that effect of the confinement on quark masses included by taking effective quark masses play a positive role and are found to improve the results further. The present $\chi$CQM results are able to give a qualitative and quantitative description of the results and further endorse the earlier
conclusion that constituent quarks and weakly interacting Goldstone bosons provide
the appropriate degree of freedom in the nonperturbative regime of QCD. The future experiments for other $J^P=\frac{3}{2}^+$ baryons will not only provide a direct method to determine the sea quark contributions  to the magnetic moments but also impose important constraints on the $\chi$CQM parameters.

{\bf ACKNOWLEDGMENTS}\\
H.D. would like to thank Department of Science and Technology (Ref No. SB/S2/HEP-004/2013), Government of India, for financial support. A.G would like to thank Harish Chandra Research Institute, Allahabad for providing the infrastructural facilities.


\begin{thebibliography}{99}

\bibitem{emc} J. Ashman {\it et al.} (EMC Collaboration), Phys.
Lett. B {\bf 206}, 364 (1988); J. Ashman { et al.} (EMC Collaboration), Nucl. Phys. B {\bf 328}, 1 (1989).

\bibitem{smc} B. Adeva {\it et al.} (SMC Collaboration), Phys. Rev.
D {\bf 58}, 112001 (1998).

\bibitem{adams} P.L. Anthony {\it et al.} (E142 Collaboration), Phys. Rev.
Lett. {\bf 71}, 959 (1993); K. Abe {\it et al.} (E143
Collaboration), Phys. Rev. Lett. {\bf 76}, 587 (1996);  K. Abe
{\it et al.} (E154 Collaboration), Phys. Rev. Lett. {\bf 79}, 26
(1997); P. Adams {\it et al.}, Phys. Rev. D {\bf 56}, 5330
(1997).

\bibitem{hermes} A. Airapetian {\it et al.} (HERMES Collaboration),
Phys. Rev. D {\bf 71}, 012003 (2005).

\bibitem{pdg}  K.A. Olive {\it et al.} (Particle Data Group), Chin. Phys. C, {\bf 38}, 090001 (2014).

\bibitem{nqm} K. Hikasa {\it et al.} (Particle Data Group), Phys. Rev. {\bf D} 45, S1 (1992) [Erratum-ibid. {\bf D} 46, 5210 (1992)].

\bibitem{rlpm} S.N. Jena and D.P. Rath, Phys. Rev. D {\bf 34}, 196
(1986).

\bibitem{rqm} F. Schlumpf, Phys. Rev. D {\bf 48}, 4478 (1993);
G. Ramalho, K. Tsushima, and F. Gross, Phys. Rev. D {\bf 80}, 033004
(2009).

\bibitem{pha} P. Ha, Phys. Rev. D {\bf 58}, 113003 (1998);
C.S. An, Q.B. Li, D.O. Riska, and B.S. Zou, Phys. Rev. C {\bf 74},
055205 (2006).

\bibitem{slaughter} M.D.
Slaughter, Phys. Rev. C {\bf 82}, 015208 (2010); M.D.
Slaughter, Phys. Rev. D {\bf 84}, 071303 (2011).

\bibitem{johan} J. Linde, T. Ohlsson, and H. Snellman, Phys. Rev.
D {\bf 57}, 452 (1998); J. Linde, T. Ohlsson, and H. Snellman, Phys. Rev.
D {\bf 57}, 5916 (1998).

\bibitem{hdmagnetic} H. Dahiya and M. Gupta, Phys. Rev. D {\bf
66}, 051501(R) (2002).

\bibitem{kerbikov} B.O. Kerbikov and Y.A. Simonov, Phys. Rev. D
{\bf 62}, 093016 (2000).

\bibitem{effmass} B.S. Bains and R.C. Verma, Phys. Rev D {\bf 66},
114008 (2002); R. Dhir and R.C. Verma, Eur. Phys. J. A {\bf 42}, 243 (2009).

\bibitem{aliev} T.M. Aliev, A. Ozpineci, and M. Savci, Phys. Rev. D {\bf 62},
053012 (2000).

\bibitem{zhu} F.X. Lee, Phys. Rev. {\bf D} 57, 1801 (1998); S.L. Zhu, W.Y.P. Hwang, and Z.S.P. Yang, Phys.
Rev. D {\bf 57}, 1527 (1998); A. Iqubal, M. Dey, and J. Dey, Phys.
Lett. B {\bf 477}, 125 (2000).


\bibitem{schwesinger} B. Schwesinger and H. Weigel, Nucl. Phys. A
{\bf 540}, 461 (1992); Y. Oh,  Phys. Rev. D {\bf 75}, 074002 (2007).

\bibitem{ledwig} T. Ledwig, A. Silva, and  M. Vanderhaeghen, Phys. Rev D
{\bf 79} 094025 (2009).

\bibitem{kim} H-C. Kim, M. Praszalowicz,  and K. Goeke, Phys. Rev. {\bf D} 57, 2859 (1998); G.S. Yang, H-C. Kim, M. Praszalowicz,  and K. Goeke, Phys. Rev. {\bf D} 70, 114002 (2004).


\bibitem{mendieta} R. Flores-Mendieta, Phys. Rev. D {\bf 80}, 094014 (2009);  L.S. Geng, J.M. Camalich, and M.J.V. Vacas, Phys. Rev D {\bf
80}, 034027 (2009).


\bibitem{boinepalli} S. Boinepalli, D.B. Leinweber, P.J. Moran,
A.G. Williams, J.M. Zanotti, and J.B. Zhang, Phys. Rev. {\bf D}
80, 054505 (2009); C. Aubin, K. Orginos, V. Pascalutsa, and M.
Vanderhaeghen, Phys. Rev. D {\bf 79}, 051502 (2009);
P.E. Shanahan {\it et al.} (CSSM and QCDSF/UKQCD Collaborations), Phys. Rev. D {\bf 89}, 074511 (2014).

\bibitem{lattice2} F.X. Lee, R. Kelly, L. Zhou and W. Wilcox, Phys. Lett. {\bf B} 627, 71 (2005).

\bibitem{hong-bagmodel} S-T. Hong, Phys. Rev. {\bf D} 76, 094029 (2007).

\bibitem{large-nc} E.E. Jenkins and A.V. Manohar, Phys. Lett. {\bf B} 335, 452 (1994); M.A. Luty, J. March-Russell, and M.J. White, Phys. Rev. {\bf D} 51, 2332 (1995); A.J. Buchmann, J.A. Hester, and R.F. Lebed, Phys. Rev. {\bf D} 66, 056002 (2002); A.J. Buchmann and R.F. Lebed, Phys. Rev. D {\bf 67},
016002 (2003).

\bibitem{heavy-chi-pert} E.E. Jenkins and A.V. Manohar, Phys. Lett. {\bf B} 255, 558 (1991).

\bibitem{buchmann} A.J. Buchmann, E. Hernandez, and A. Faessler,
Nucl. Phys. A {\bf 569}, 661 (1994); G. Wagner, A.J. Buchmann, and A. Faessler, Phys.
Lett. B {\bf 359}, 288 (1995);  A.J. Buchmann, E. Hernandez, and A. Faessler,
Phys. Rev. C {\bf 55}, 448 (1997); G. Wagner, A.J. Buchmann, and A. Faessler, Phys.
Rev. C {\bf 58}, 3666 (1998); G. Wagner, A.J. Buchmann, and A. Faessler, J. Phys.
G {\bf 26}, 267 (2000); A.J. Buchmann and E.M. Henley, Phys. Rev. {\bf C
63}, 015202 (2000);  A.J. Buchmann, Phys. Rev. Lett. {\bf 93}, 212301
(2004).



\bibitem{manohar} S. Weinberg, Physica A {\bf 96}, 327 (1979); A.
Manohar and H. Georgi, Nucl. Phys. B {\bf 234}, 189 (1984).

\bibitem{eichten} E.J. Eichten, I. Hinchliffe, and C. Quigg, Phys.
Rev. D {\bf 45}, 2269 (1992).

\bibitem{cheng} T.P. Cheng and L.F. Li, Phys. Rev. Lett. {\bf
74}, 2872 (1995); T.P. Cheng and L.F. Li, Phys. Rev. D {\bf 57}, 344 (1998); T.P. Cheng and L.F. Li, Phys. Rev. Lett.
{\bf 80}, 2789 (1998).

\bibitem{song} X. Song, J.S. McCarthy, and H.J. Weber, Phys. Rev.
D {\bf 55}, 2624 (1997); X. Song, Phys. Rev. D {\bf 57}, 4114
(1998).

\bibitem{hd} H. Dahiya and M. Gupta, Phys. Rev. D {\bf 64}, 014013
(2001); H. Dahiya and M. Gupta, Int. Jol.
of Mod. Phys. A, Vol. 19, No. 29, 5027 (2004); H. Dahiya, M. Gupta
and J.M.S. Rana, Int. Jol. of Mod. Phys. A, Vol. 21, No. 21, 4255
(2006); H. Dahiya and M. Randhawa, Phys. Rev. D {\bf 90}, 074001 (2014).

\bibitem{hds} H. Dahiya and M. Gupta, Eur. Phys. J. C {\bf 52}, 571
(2007); H. Dahiya and M. Gupta, Phys. Rev. D {\bf 78}, 014001 (2008).

\bibitem{dgg} A. De Rujula, H. Georgi, and S.L. Glashow, Phys. Rev.
D {\bf 12}, 147 (1975).

\bibitem{isgur} N. Isgur, G. Karl, and R. Koniuk, Phys. Rev. Lett.
{\bf 41}, 1269 (1978); R. Koniuk and N. Isgur, Phys. Rev. D {\bf
21}, 1868 (1980); N. Isgur and G. Karl, Phys. Rev. D {\bf 21}, 3175
(1980); N. Isgur, G. Karl, J. Soffer, Phys. Rev. D {\bf 35}, 1665 (1987).

\bibitem{yaouanc} A. Le Yaouanc, L. Oliver, O. Pene and J.C. Raynal,
 Phys. Rev. D {\bf 12},  2137 (1975); {\it ibid.} {\bf 15}, 844 (1977).

\bibitem{mgupta} M. Gupta, S.K. Sood, and A.N. Mitra, Phys. Rev. D {\bf 16}, 216 (1977); M. Gupta and A.N. Mitra, Phys. Rev. D {\bf 18}, 1585 (1978); M. Gupta, S.K. Sood, and A.N. Mitra, Phys. Rev. D {\bf 19}, 104 (1979); M. Gupta and N. Kaur,
Phys. Rev. D {\bf 28}, 534 (1983); P.N. Pandit, M.P. Khanna, and M. Gupta,
J. Phys. G {\bf 11}, 683 (1985).


\bibitem{nsweak} N. Sharma, H. Dahiya, P.K. Chatley, and M. Gupta,
 Phys. Rev. D {\bf 79}, 077503 (2009); N. Sharma, H. Dahiya and P.K.
Chatley, Eur. Phys. J. A {\bf 44}, 125 (2010), N. Sharma and H. Dahiya, Phys. Rev. D {\bf 81},  114003 (2010).


\bibitem{nres} N. Sharma, A.M. Torres, K.P. Khemchandani, and H. Dahiya,
Eur. Jol. Phys. A {\bf 49}, 11 (2013).

\bibitem{torres} A.M. Torres, K.P. Khemchandani, N. Sharma, and H. Dahiya, Eur. Jol. Phys. A {\bf 48}, 185 (2012).

\bibitem{charge-radii}	N. Sharma and H. Dahiya,
Pramana, {\bf 81}, 449 (2013);	N. Sharma and H. Dahiya, Pramana, {\bf 80}, 237 (2013).

\bibitem{hdcharm} H. Dahiya and M. Gupta, Phys. Rev. D {\bf 67},
074001 (2003); N. Sharma, H. Dahiya, P.K. Chatley, and M. Gupta
Phys. Rev. D {\bf 81}, 073001 (2010); N. Sharma and H. Dahiya, Int. Jol. of Mod. Phys. A, Vol. 28, No. 14, 1350052 (2013).

\bibitem{effm} Ikuo S. Sogami and Noboru Oh'yamaguchi, Phys. Rev. Lett.
{\bf 54}, 2295 (1985); Kuang-Ta Chao, Phys. Rev. D {\bf 41}, 920 (1990); M. Gupta, J. Phys. G {\bf 16}, L213 (1990).


\bibitem{e866} E866/NuSea Collaboration, E.A. Hawker {\it et al.},
 Phys. Rev. Lett. {\bf 80}, 3715 (1998); J.C. Peng {\it et al.},
 Phys. Rev. D {\bf 58}, 092004 (1998); R.S. Towell {\it et al.},
Phys. Rev. D {\bf 64}, 052002 (2001).

\bibitem{GSR}  K. Gottfried, Phys. Rev. Lett. {\bf 18}, 1174 (1967).

\bibitem{mu1}  J. Franklin, Phys. Rev. {\bf  172}, 1807 (1968);
N. Isgur, Acta Physica Polonica B {\bf 8}, 1081 (1977);
G.E. Brown, M. Rho and V. Vento,
Phys. Lett. {\bf  97B}, 423 (1980).

\bibitem{cgsr} S. Coleman and S.L. Glashow, Phys. Rev. Lett. {\bf
6}, 423(1961).

\bibitem{prl-deltaplus} M. Kotulla {\it et al.} Phys. Rev. Lett.
{\bf 89}, 272001 (2002).



\end{thebibliography}
\end{document}